# Nanoplasmonic Tweezers Visualize Protein p53 Suppressing Unzipping of Single DNA-Hairpins

*Abhay Kotnala and Reuven Gordon\**

*Department of Electrical Engineering, University of Victoria, Victoria, British Columbia V8W 3P6, Canada*

ABSTRACT: Here we report on the use of double-nanohole (DNH) optical tweezers as a label-free and free-solution single-molecule probe for protein–DNA interactions. Using this approach, we demonstrate the unzipping of individual 10 base pair DNA-hairpins, and quantify how tumour suppressor p53 protein delays the unzipping. From the Arrhenius behaviour, we find the energy barrier to unzipping introduced by p53 to be $2 \times 10^{-20}$ J, whereas cys135ser mutant p53 does not show suppression of unzipping, gives clues to its functional inability to suppress tumour growth. This transformative approach to single molecule analysis allows for ultra-sensitive detection and quantification of protein–DNA interactions to revolutionize the fight against genetic diseases.

Optical tweezers have been used to study protein–DNA interactions, critical for maintaining genetic functionality and integrity, at the single molecule level.[1-4] Although optical tweezers offer forces in the pN-nm range (around kT), relevant for the study of

protein–DNA interactions, they suffer from the use of tethering, fluorescent labeling and being limited to long DNA chains (~10 kbps).[5-9] The tether is required because conventional tweezers are not able to hold on to single molecules without using damaging laser powers.[10] Tethering also limits tweezer studies to relatively long DNA strands (~10 kbps),[6] where the localization of sites of interest is difficult. Fluorescent labels are used to monitor dynamic processes, such as protein binding[11] and unzipping[7]. Fluorescence labels change the molecule (except for autofluorescence), heat the molecule, suffer from photo bleaching,[12] have a poor signal-to-noise ratio (SNR)[13] and are limited to millisecond time-scales[14]. Adding labels and tethers modifies the natural state of the DNA/protein[11] and restricts free motion, as well as adding cost and complexity. To overcome these limitations of conventional optical tweezers, an ideal approach would be label-free, free-solution and work at the single DNA level with a small number of base-pairs. This approach should also be simple, low-cost, scalable and use low laser power.

Here we demonstrate that the DNH laser tweezer, a nanoplasmonic structure, can overcome the limitations of conventional optical tweezers in the study of protein–DNA interactions for short DNA chains, without the need for tethering and without the need for fluorescent labels. The DNH tweezer approach uses low optical powers and a conductive gold film, so there is negligible heating (estimated to be ~0.1K).[15] The scattering signal observed in trapping and unzipping is extraordinarily high, with laser transmission changes of around 10% being typical, even for nanoparticles/molecules in the single nanometer range. We demonstrate that the DNH tweezer can easily trap and unzip a 10 bp DNA hairpin. We further show that tumor suppressor protein p53 retards the unzipping, from which we quantify the p53 unzipping suppression energy using Arrhenius scaling. Mutant p53, on the other hand, does not suppress unzipping, which may explain its ineffectiveness in tumour suppression. This shows, we believe for the first time, the direct role of p53 in suppressing DNA unzipping.

The DNH optical tweezer uses simple inverted microscope geometry as shown in Figure 1. A 820 nm laser beam (Sacher Lasertechnik) is focused onto the DNH using a 100× oil immersion objective (1.25 numerical aperture). The transmission through the DNH is measured using a 50 MHz bandwidth avalanche photodiode (APD) and the transmission is maximized with a half wave plate by aligning the laser polarization along the DNH cusp axis. Even for low laser powers (between 1-3 mW at the DNH), an optical density filter (OD 1) is required to prevent saturation of the APD. The DNH focuses the laser power to the nanometric gap between the cusps, favourable for trapping nanoparticles and biomolecules with low incident laser power.[16,17] Even for particles at the nanometer scale, the DNH tweezer provides trapping efficiency in the range of pN-nm,[18] which is the regime of the thermal energy at room temperature (kT) relevant to the study of natural biomolecular interactions. The biochip (Figure 1, zoomed region) consists of DNHs fabricated on commercially available 100 nm thick gold test slides (EMF corp.) using a focussed ion beam. For trapping of 10 bp hairpins, the cusp spacing was fabricated to be ~ 10 nm (Fig. 1 inset). Note that this is significantly smaller than used in past works,[19,20] which is a key enabling feature of the present study.

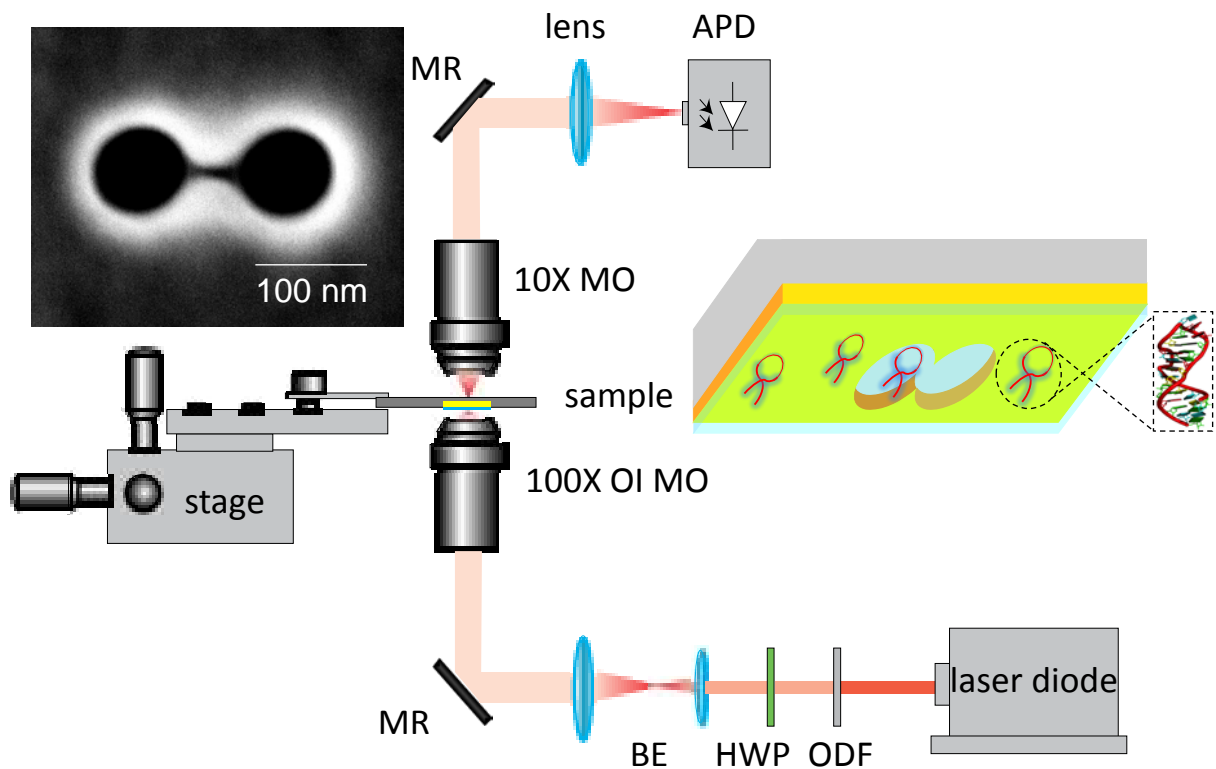

**Figure 1.** Schematic of DNH laser tweezer trapping and unzipping DNA hairpins. The circular inset shows gold sample with DNH and DNA hairpins suspended in phosphate buffer solution. The rectangular inset is a scanning electron microscope (SEM) image of the DNH fabricated using focused ion beam (FIB). Abbreviation used: ODF = optical density filter; HWP = half wave plate; BE = beam expander; MR: mirror; OIMO: oil immersion microscope objective; APD = avalanche photodiode.

Figure 2 shows the detected laser transmission at the APD for the DNH tweezer trapping events for 20 base DNA strands. Figure 2a shows the trapping event, seen with a discrete jump in the APD voltage, for a single-stranded DNA that does not hairpin. The discrete jump in transmission is due to dielectric loading of the DNH.[19] Figure 2b shows the trapping event for a single-strand that forms a hairpin. It is clearly seen that there is a double-step in Figure 2b. We attribute the double step to the unzipping of the 20 base (10 bp) hairpin.

(Note that the unzipped single stranded DNA is only 7 ~ nm long). This was observed consistently for hairpin structures, but never for single-stranded DNAs that do not hairpin. The DNH tweezer energetically favours unzipping of the hairpin because the elongated DNA has a larger polarizability than one that is zipped up, and the polarizability determines the interaction energy with the laser field. Thus we have shown the ability of the DNH tweezers to trap 20 base DNA and unzip the hairpin structure over a typical timescale of 0.1 s. Figure 2c shows a simple energy reaction diagram showing the trapping and unzipping of the hairpin DNA. The hairpin DNA is initially trapped with energy of ~ 10 kT, as suggested by Ashkin for stable trapping[21] and followed by unzipping requiring approximately the same energy. We estimate the energy change by the size of the transmission change through the aperture. This is evident from the similar change in the transmission intensity during trapping and unzipping.

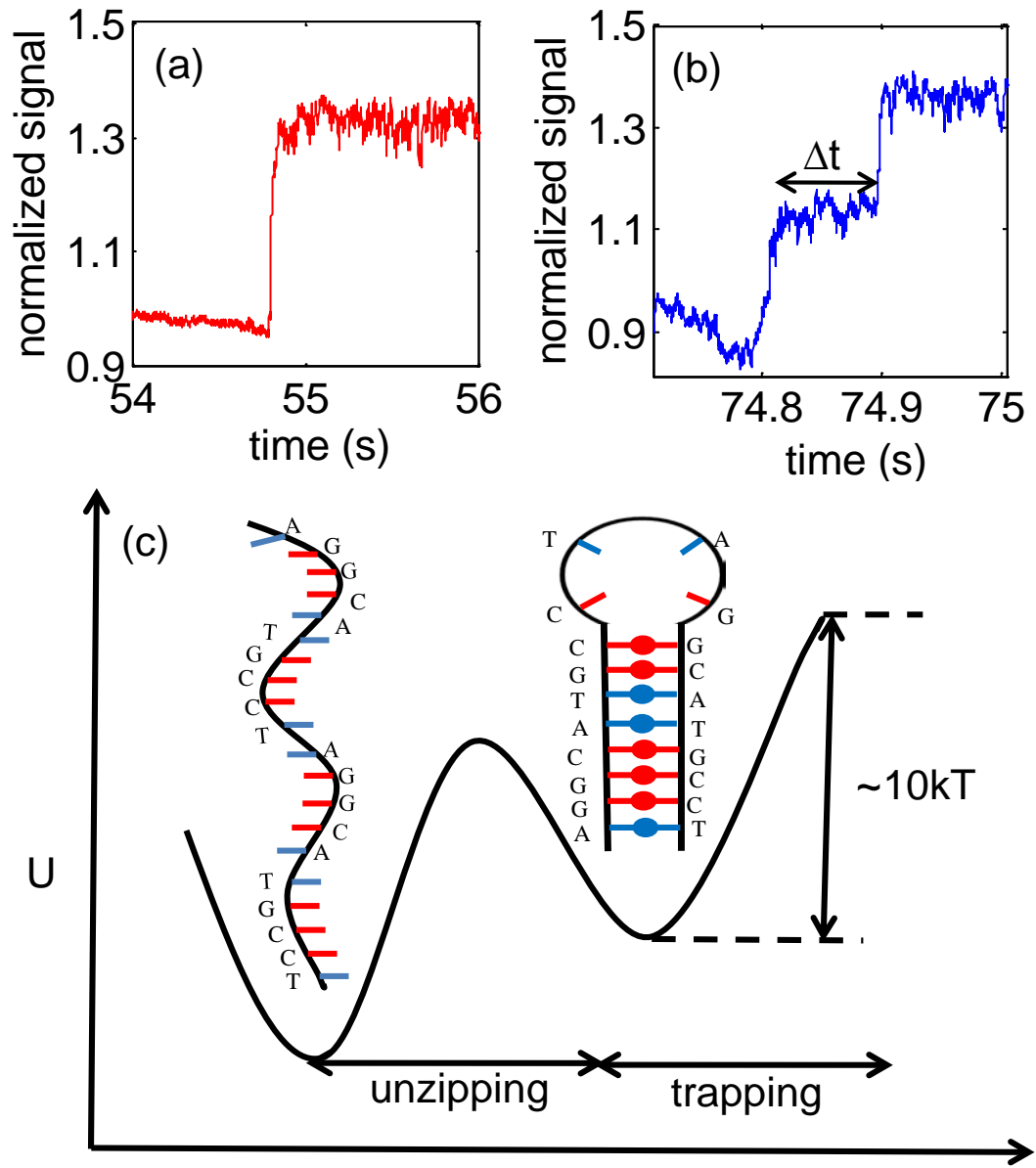

**Figure 2.** Trapping and unzipping 20 base DNA strands. a) Single strand DNA trapping event with no intermediate step b) A hairpin DNA trap event showing the unzipping with an intermediate step of ~ 0.1s. c) Energy reaction diagram of trapping and unzipping of DNA hairpin. k: Boltzmann constant, T: Temperature, U: Energy

DNA binding proteins can stabilize or destabilize the DNA structure, which can impact unzipping.[2] In the present case, we study the tumour suppressor p53 protein – DNA interaction, for which the suppression of unzipping has not been established, but fluorescence anisotropy works have shown the binding strength[22].

Figure 3a shows the trapping signal of the p53 wild type protein– DNA complex with a long unzipping time (Δt). The increased unzipping time is associated with the strong binding of p53 with the consensus DNA hairpin structure [23], critical for the biological activity of p53.[24] The unzipping time (Δt) can be used to quantify the unzipping suppression energy of p53 protein–DNA interaction.[25] The cumulative probability plot shown in Figure 3b shows the unzipping time Δt always greater than 1s for wild type p53–DNA complex as compared to that of DNA for a given probability range.

The energy reaction diagram for the protein–DNA complex, as shown Figure. 3c, is similar to that of hairpin DNA except for an increase in the energy barrier (ΔU) between the trapped and unzipped state. The increase in energy barrier ΔU results in longer unzipping time (Δt) and is a measure of the unzipping suppression energy ΔU. Therefore using the Arrhenius behavior ΔU is given by

$$\Delta U = -kT \ln \frac{t_{p53}}{t_{DNA}}$$

Where, $t_{p53}$ and $t_{DNA}$ is the mean unzipping time obtained from a log-normal distribution fit to unzipping time Δt of p53– DNA and DNA respectively. The mean values obtained are $t_{p53} = 7.9$ s and $t_{DNA} = 0.1$ s

Using the above values we find the binding energy $\Delta G = 2.0 \times 10^{-20}$ J. While p53 is well-known to suppress tumors, as far as we can tell no works have suggested that p53 suppresses DNA unzipping, let alone quantified the energy for this suppression.[26,27] Recent electron microscopy studies have presented a tetramer structure for p53 that encapsulates the DNA;

however, the function with respect to DNA unzipping of this structure has not been addressed.[28,29] The energy of unzipping suppression is expected to be less than, but of the same order of magnitude as the maximal binding energy for p53, which is found to be $6.9 \times 10^{-20}$ J by fluorescence anisotropy studies.[27] For the particular DNA sequence we are using, there is a 2 bp mismatch from the optimal p53 binding structure, and so the binding energy is expected to be $6.7 \times 10^{-20}$ J, using a previously proposed formulation for binding energy[27].

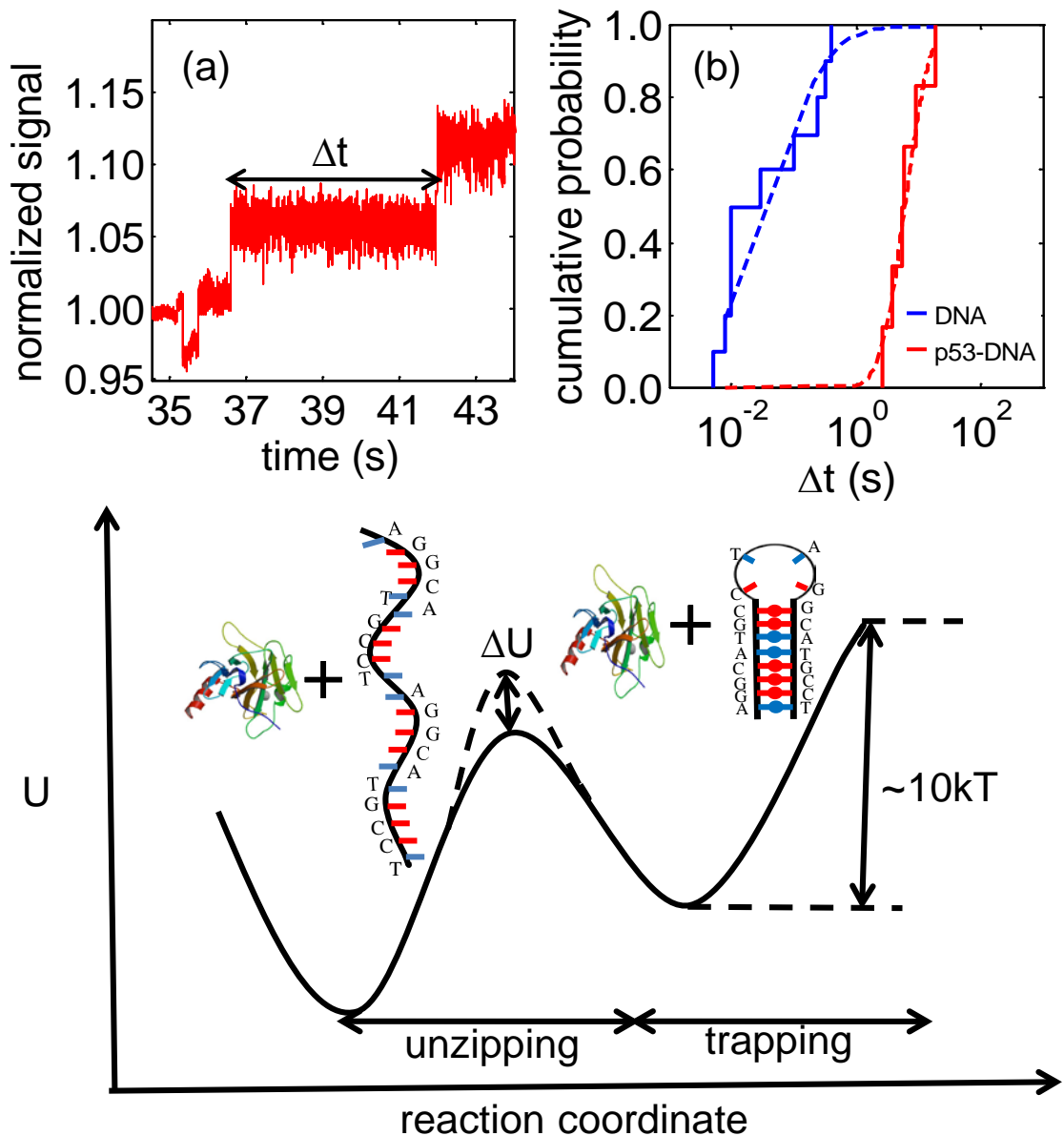

**Figure 3.** Suppression of DNA hairpin unzipping by tumour suppressor protein p53. a) The wild type p53 suppresses the unzipping of the DNA hairpin for a delay of ~10 seconds. b) Comparison of cumulative probability of unzipping time Δt for p53-DNA complex and DNA alone. c) Energy reaction diagram showing increased energy barrier ΔU equivalent to the binding energy Δ*G* of p53 and DNA.

To understand the interaction of the hairpin DNA with a p53 mutant we also trap the p53 mutant-DNA complex. Figure 4a shows the trapping signal for p53 mutant–DNA complex, showing no appreciable difference in the unzipping time Δt. This is interesting because the single point mutation of cys135ser results in only partial loss of DNA binding [24] but here we show that it completely loses the ability to suppress the unzipping of the hairpin DNA. The cumulative probability distribution of the unzipping time Δt shows an overlap with that of DNA as shown in Figure 4b. Thus the DNH tweezers also shows the ability to distinguish between the interactive behaviour of the p53 wild type and its mutant protein with hairpin DNA; that is, it shows the specificity required for a good sensor/detector.

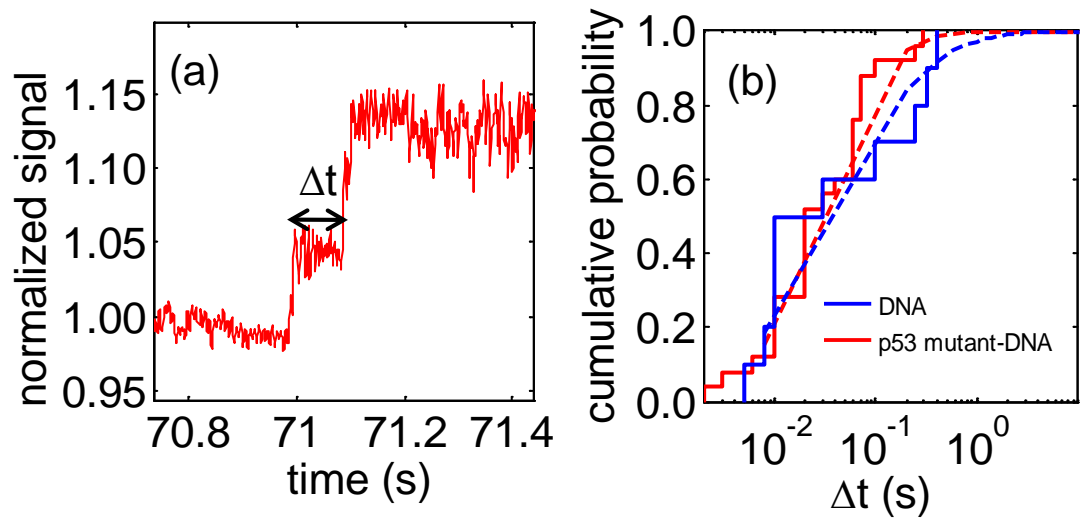

**Figure 4.** Unzipping DNA hairpin and influence of mutant p53. a) The mutant p53 is incapable to suppress the unzipping of the DNA hairpin even though there is partial loss in binding activity (b) Cumulative probability of unzip time Δt for mutant p53– DNA and hairpin DNA.

To ensure that we were not trapping p53 alone in the above measurements, we trapped p53 wild type and its mutant individually without the DNA. Typical events are shown in Figure 5a and 5b. Both the trapping events look almost identical with small optical scattering and unstable behavior. These events have a much smaller step than from the protein − DNA complex and from the DNA alone. The nearly identical behavior for the mutant and the wild-type is because of minimal structural difference between the two proteins as has been illustrated for single point cy238ser mutant numerically.[30]

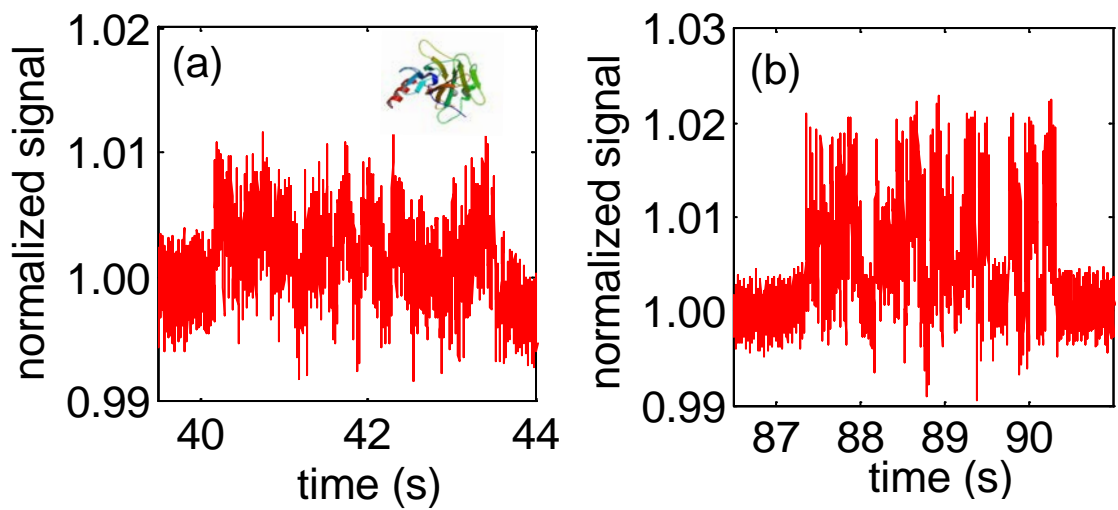

**Figure 5.** Weak optical trapping of p53 proteins. Optical trapping of p53 (a) wild-type and (b) mutant alone. Inset shows p53 protein structure.

In summary a label-free, free solution, sensitive and low cost DNH optical tweezer was used to show the unzipping of the hairpin DNA structure and its interaction with the tumour suppressor p53 protein. We showed the suppression of unzipping by wild type p53 protein due to strong binding with the consensus hairpin DNA and evaluate the unzipping suppression energy based on Arrhenius behaviour to be $2.0 \times 10^{-20}$ J. The mutant (cys135ser) shows negligible impact on the unzipping of the DNA even though there is only partial loss in the binding activity, which may explain its ineffectiveness in suppression tumour development. Thus DNH tweezers show the ability to understand the dynamics of small DNA fragments and the capability to distinguish the impact of a normal protein from its mutant on their behaviour.

We believe that this capability will have a transformative impact on biosensors. For example, we can distinguish mutant from wild-type species. It also shows great promise for drug discovery; for example, for the p53 case shown, the influence of small molecules that allow the mutant form to function normally would be of great interest. Finally, our work represents an almost ideal research tool to better understand how protein – DNA interactions (and other similarly sized molecules) in real time, at the single molecule level, in free-solution and in a label-free way.

## Methods

**Preparation of DNA solution.** The 20 base hairpin DNA sequence, 5'- AGG CAT GCC TAG GCA TGC CT -3' and a single strand DNA sequence, 5'-GGG CGG GGA GGG GGA AGG GA -3' were used (Integrated DNA Technologies). The DNA fragments were re-suspended in the phosphate buffer solution (PBS) of pH 7.5 to 0.02% w/v concentration. The p53 human recombinant full length protein (Cedarlane Labs, CLPRO742) and its mutant cys135ser (Cedarlane Labs, CLPRO301) are produced in E.coli. The protein – DNA complex solution was prepared with 1:1 ratio of protein and DNA by volume. This was done for both p53 wild type and its mutant.

**Nanofabrication of the DNH.** The DNH with separation of ~ 10 nm between the cusps were fabricated using Hitachi FB-2100 focused ion beam system. The DNH were fabricated on gold test slides (EMF Corporation) with 100nm thick gold layer with a 5 nm Ti adhesion layer on a glass substrate. A gallium ion beam of accelerating voltage 40kV and current 0.001nA was used to mill the gold at a magnification of 80K. The beam of dwell time 10μs with 30 passes fabricates the DNH with ~ 10 nm cusp separation.

AUTHOR INFORMATION

**Corresponding Author**

*E-mail: rgordon@uvic.ca

**Notes**

The authors declare no competing financial interest.

ACKNOWLEDGMENTS

This work was supported by the NSERC (Canada) Discovery Grants program.

REFERENCES

(1) Moffitt, J. R.; Chemla, Y. R.; Smith, S. B.; Bustamante, C. *Annu. Rev. Biochem.* **2008**, *1,* 205-228.

(2) Koch, S. J.; Shundrovsky, A.; Jantzen, B. C.; Wang, M. D. *Biophys. J.* **2002**, *2,* 1098-1105.

(3) Shokri,L; Rouzina.I; Williams.M.C, *Physical Biology* **2009**, *2,* 025002.

(4) Shaevitz, J. W.; Abbondanzieri, E. A.; Landick, R.; Block, S. M. *Nature* **2003**, *6967,* 684-687.

(5) Heller, I.; Hoekstra, T. P.; King, G. A.; Peterman, E. J. G.; Wuite, G. J. L. *Chem. Rev.* **2014**.

(6) Farge, G.; Laurens, N.; Broekmans, O. D.; van, d. W.; Dekker, L. C. M.; Gaspari, M.; Gustafsson, C. M.; Peterman, E. J. G.; Falkenberg, M.; Wuite, G. J. L. *Nat. Commun.* **2012**, 1013.

(7) Bianco, P. R.; Brewer, L. R.; Corzett, M.; Balhorn, R.; Yeh, Y.; Kowalczykowski, S. C.; Baskin, R. J. *Nature* **2001**, *6818,* 374-378.

(8) Neuman, K. C.; Nagy, A. *Nat. Meth.* **2008**, *6,* 491-505.

(9) Chaurasiya, K. R.; Paramanathan, T.; McCauley, M. J.; Williams, M. C. *Phys. Life Rev.* **2010**, *3,* 299-341.

(10) Peterman, E.; Gittes, F.; Schmidt, C. *Biophys. J.* **2003**, *2,* 1308-1316.

(11) Heller, I.; Sitters, G.; Broekmans, O. D.; Farge, G.; Menges, C.; Wende, W.; Hell, S. W.; Peterman, E. J. G.; Wuite, G. J. L. *Nat. Meth.* **2013**, *9,* 910-916.

(12) Van Dijk, M.; Kapitein, L. C.; Van Mameren, J.; Schmidt, C. F.; Peterman, E. *J Phys Chem B* **2004**, *20,* 6479-6484.

(13) Waters, J. C. *J. Cell Biol.* **2009**, *7,* 1135-1148.

(14) Hohng, S.; Zhou, R.; Nahas, M. K.; Yu, J.; Schulten, K.; Lilley, D. M. J.; Ha, T. *Science* **2007**, *5848,* 279-283.

(15) Melentiev, P. N.; Afanasiev, A. E.; Kuzin, A. A.; Baturin, A. S.; Balykin, V. I. *Opt. Express* **2013**, *12,* 13896-13905.

(16) Pang, Y.; Gordon, R. *Nano Lett.* **2012**, *1,* 402-406.

(17) Juan, M. L.; Righini, M.; Quidant, R. *Nat. Photon.* **2011**, *6,* 349-356.

(18) Kotnala, A.; Gordon, R. *Nano Lett.* **2014**.

(19) Juan, M. L.; Gordon, R.; Pang, Y.; Eftekhari, F.; Quidant, R. *Nat. Phys.* **2009**, *12,* 915-919.

(20) Kotnala, A.; DePaoli, D.; Gordon, R. *Lab Chip* **2013**, *20,* 4142-4146.

(21) Ashkin, A.; Dziedzic, J. M.; Bjorkholm, J. E.; Chu, S. *Opt. Lett.* **1986**, *5,* 288-290.

(22) Veprintsev, D.; Fersht, A. *Nucleic Acids Res.* **2008**, *5,* 1589-1598.

(23) Gohler, T.; Reimann, M.; Cherny, D.; Walter, K.; Warnecke, G.; Kim, E.; Deppert, W. *J. Biol. Chem.* **2002**, *43,* 41192-41203.

(24) Buzek, J.; Latonen, L.; Kurki, S.; Peltonen, K.; Laiho, M. *Nucleic Acids Res.* **2002**, *11,* 2340-2348.

(25) Hall, M. A.; Shundrovsky, A.; Bai, L.; Fulbright, R. M.; Lis, J. T.; Wang, M. D. *Nat. Struct. Mol. Biol.* **2009**, *2,* 124-129.

(26) Chen, Y.; Zhang, X.; Dantas Machado, A. C.; Ding, Y.; Chen, Z.; Qin, P. Z.; Rohs, R.; Chen, L. *Nucleic Acids Res.* **2013**, *17,* 8368-8376.

(27) Lukman, S.; Lane, D. P.; Verma, C. S. *PloS one* **2013**, *11,* e80221.

(28) Melero, R.; Rajagopalan, S.; Lázaro, M.; Joerger, A. C.; Brandt, T.; Veprintsev, D. B.; Lasso, G.; Gil, D.; Scheres, S. H. W.; Carazo, J. M.; Fersht, A. R.; Valle, M. *Proc. Natl. Acad. Sci.* **2011**, *2,* 557-562.

(29) Cherny, D. I.; Striker, G.; Subramaniam, V.; Jett, S. D.; Paleček, E.; Jovin, T. M. *J. Mol. Biol.* **1999**, *4,* 1015-1026.

(30) Ferrone, M.; Perrone, F.; Tamborini, E.; Paneni, M. S.; Fermeglia, M.; Suardi, S.; Pastore, E.; Delia, D.; Pierotti, M. A.; Pricl, S.; Pilotti, S. *Mol. Cancer Ther.* **2006**, *6,* 1467-1473.